\documentclass[11pt]{article}
\usepackage{epsfig}
\usepackage{psfig}
\input{epsf}
\newcommand{\beq}{\begin{equation}}
\newcommand{\eeq}{\end{equation}}
\newcommand{\beqa}{\begin{eqnarray}}
\newcommand{\eeqa}{\end{eqnarray}}

\newcommand{\ve}{\varepsilon}

\newcommand{\vs}{\vspace{-0.29cm}}
\begin{document}

%%\noindent Accepted for publication in 

\begin{flushright}
{\tiny{HISKP-TH-05/08,}}
{\tiny{FZJ-IKP-TH-2005-11,}}
{\tiny{TUM-T39-05-01}}
\end{flushright}

\vspace{.4in}

\begin{center}

\bigskip

{{\Large\bf Chiral Extrapolations and the Covariant Small Scale Expansion}\footnote{This
research is part of the EU Integrated Infrastructure Initiative Hadron Physics Project
under contract number RII3-CT-2004-506078. Work supported in part by DFG (SFB/TR 16,
``Subnuclear Structure of Matter'') and BMBF.}}

\end{center}

\vspace{.3in}

\begin{center}
{\large 
V\'eronique Bernard$^\star$\footnote{email: bernard@lpt6.u-strasbg.fr},
Thomas R. Hemmert$^\dagger$\footnote{email: themmert@physik.tu-muenchen.de},
Ulf-G. Mei{\ss}ner$^\ddagger$$^\ast$\footnote{email: meissner@itkp.uni-bonn.de}
}

\vspace{1cm}

$^\star${\it Universit\'e Louis Pasteur, Laboratoire de Physique
            Th\'eorique\\ 3-5, rue de l'Universit\'e,
            F--67084 Strasbourg, France}

\bigskip

$^\dagger${\it Physik-Department, Theoretische Physik T39\\
    TU M{\"u}nchen, D-85747 Garching, Germany}\\

\bigskip

$^\ddagger${\it Universit\"at Bonn,
Helmholtz--Institut f\"ur Strahlen-- und Kernphysik (Theorie)\\
Nu{\ss}allee 14-16,
D-53115 Bonn, Germany}

\bigskip

$^\ast${\it Forschungszentrum J\"ulich, Institut f\"ur Kernphysik 
(Theorie)\\ D-52425 J\"ulich, Germany}

\bigskip

\bigskip

\end{center}

\vspace{.4in}

\thispagestyle{empty} 

\begin{abstract}\noindent 
We calculate the nucleon and the delta mass to fourth order 
in a covariant formulation of the small scale expansion. 
We analyze lattice data from the MILC collaboration and demonstrate that the 
available lattice data combined with our knowledge
 of the physical values for the nucleon and delta masses lead to consistent 
chiral extrapolation functions for both observables up to fairly large
 pion masses. This holds in particular for very recent data on the
delta mass from the QCDSF collaboration.
The resulting pion-nucleon sigma term is $\sigma_{\pi N}=48.9$ MeV. This 
first quantitative analysis of the quark-mass dependence of the structure of
the $\Delta(1232)$ in full QCD within chiral effective field theory
 suggests that (the real part of) the 
nucleon-delta mass-splitting in the chiral limit, $\Delta_0=0.33\,$GeV, 
is slightly larger than at the physical point. Further analysis of 
{\em simultaneous} fits to nucleon and delta lattice data are needed for a 
precision determination of the properties of the first 
excited state of the nucleon.
\end{abstract}

\vfill

\pagebreak

%%%%%%%%%%%%%%%%%%%%%%%%%%%%%%%%%%%%%%%%%%%%%%%%%%%%%%%%%%%%%%%%%%%%%%%
\noindent {\bf 1.} The $\Delta (1232)$  is the most important baryon
resonance. It is almost degenerate in mass with the nucleon
and couples strongly to pions, nucleons and photons. It was therefore
argued early that spin-3/2 (decuplet) states should be included in baryon
chiral perturbation theory \cite{JMdel}, which is the low-energy effective 
field theory of the Standard Model. In Ref.\cite{JMdel} and 
subsequent works use was made of the heavy baryon approach, which treats 
the baryons as static sources. However, due to the fact that the nucleon-delta
mass  splitting $\Delta=m_\Delta-m_N$ stays 
finite in the chiral limit and thus introduces an additional low energy scale, 
in chiral effective field theories (ChEFT) special
 care has to be taken about the decoupling of resonances in the chiral limit \cite{GZ}. 
This  requirement can be systematized
by counting the nucleon-delta mass splitting as an additional small 
parameter \cite{HHK}, which ensures that at each order in the chiral expansion
enough counterterms are present to 
guarantee decoupling and renormalization \cite{BFHM}. 
The corresponding power counting was called the  ``small scale
expansion'' (SSE) \cite{HHK}. The heavy baryon approach has 
been successfully applied to a
variety of processes, for reviews see \cite{BKMrev,UGM}, and a status report
on chiral effective field theories with deltas is given in \cite{TRH}.
More recently, it was realized that for certain considerations/processes a 
Lorentz-invariant formulation of baryon chiral perturbation
theory is advantageous. A particularly elegant scheme to perform covariant
calculations is the so--called ``infrared regularization'' (IR) of \cite{BL}.
The renormalization of relativistic baryon chiral perturbation theory has been 
discussed in detail in \cite{FGJS}. In Ref.\cite{SSEIR} we gave
a consistent extension of the infrared regularization method in the presence
of  spin-3/2 fields. It was in particular shown that in the covariant formulation of 
the SSE the contribution of the 
non-propagating spin-1/2  components of the Rarita-Schwinger field can be
completely absorbed in the polynomial terms stemming from the most general
effective chiral Lagrangian, analogous to the situation in the non-relativistic
SSE \cite{HHK}. In this letter, we apply the covariant 
SSE formalism to the
nucleon and the delta mass\footnote{For the substantial literature on this
  topic analysed within the heavy baryon approach 
we refer to the most recent work \cite{HB} and the literature cited therein. 
For related early work on the problem of chiral 
extrapolation functions for $\Delta$(1232) properties see \cite{aus}.} as well
as the $\pi N$ sigma term. 
We perform a fourth order calculation in the small parameter $\ve$, where
$\ve$ collects small external momenta, the pion
mass and the $N\Delta$ mass splitting. These explicit representations of 
$m_N$ and $m_\Delta$ serve as chiral extrapolation functions to analyze
lattice simulations for nucleon and delta masses involving dynamical 
fermions\footnote{For a discussion of the 
chiral extrapolation of nucleon and delta masses in quenched QCD see
\cite{James}.}, as for example reported by the 
CPPACS \cite{CPPACS} and JLQCD collaboration \cite{JLQCD}. 
While a detailed (combined) analysis of these and forthcoming 
data \cite{QCDSF} analogous to the work reported in Ref.\cite{MN}
certainly also needs to take into account the observed finite volume 
dependences \cite{CPPACS,JLQCD}, in this letter\footnote{Some of our 
results have been reported previously \cite{conferences}.} our aim is 
more modest. We attempt to connect the lattice data for 
nucleon and delta masses as reported by 
the MILC collaboration \cite{MILC} with the known results at the physical
point. The MILC data have the exciting feature that they cover 
unphysical pion masses as low as $M_\pi \simeq 350$~MeV on a relatively large 
lattice with $L=2.6$~fm, although there is still 
considerable discussion within the lattice QCD 
community about the technical details of the employed (staggered) 
fermion actions, see e.g. \cite{Karl}. Clearly, such extrapolation functions based on 
chiral perturbation theory cease to make sense at too large values of the quark (pion)
masses, but as we will demonstrate later, we can nicely capture the trend of
these data\footnote{The bulk of the MILC data have been obtained in simulation 
runs with three active flavours, while our chiral  extrapolation functions
apply to a scenario with only two light quark flavours. However, at the 
(still relatively large) quark masses 
studied in Ref.\cite{MILC}, corresponding to pion masses $M_\pi > 350$ MeV, 
the MILC collaboration reported no significant differences 
between their two and three flavour runs for the observables considered here. 
Note that they published a single two-flavor run at one quark mass that
showed no sizable deviations from the three-flavor runs in \cite{MILC}.
We will therefore treat the MILC data as if they constituted 
dynamical two flavor results. This is further corroborated by our analyses of
the nucleon mass in SU(2) and SU(3) heavy baryon CHPT - no large effects were
found when going from two to three flavors, see \cite{BHMcut} and \cite{FMS},
respectively.}. More lattice results also at 
lower pion masses are clearly needed, and these should be analyzed utilizing
the (infinite volume) extrapolation functions given here.

\medskip

%%%%%%%%%%%%%%%%%%%%%%%%%%%%%%%%%%%%%%%%%%%%%%%%%%%%%%%%%%%%%%%%%%%%%%%%%%%%%%
\noindent {\bf 2.}  Our calculations are based on the effective
Lagrangian of nucleons and deltas coupled to external sources.
The various contributions to S-matrix elements and transition 
currents are organized in powers of the small parameter $\ve$,
where $\ve$ collectively denotes small pion four-momenta, the
pion mass, baryon three momenta and the nucleon-delta splitting,
$\Delta = m_\Delta - m_N$ (more precisely, the difference in the
chiral limit). The expansion in powers of $\ve$ is called the 
small scale expansion. In what follows, we will consider the
nucleon and the delta mass to fourth order in the SSE. 
Consider first the pion-nucleon Lagrangian.
The terms pertinent to the observables calculated here read (for
details, see \cite{BFHM,BKMrev,FMMS}), 
\begin{eqnarray}
{\cal L}_{\pi N}^{(1)}&=&\bar{\psi}_N\left[i\not{\!\!D}-m_0+\frac{g_A}{2}\not{u}
                         \gamma_5\right]\psi_N ~, \label{s121} \\
{\cal L}_{\pi N}^{(2)}&=&\bar{\psi}_N\left[c_1 \langle\chi_+\rangle
                         -\frac{c_2}{4 m_0^2}\left\{\langle u_\mu u_\nu
                         \rangle D^\mu D^\nu+ {\rm h.c.}\right\}
                         +\frac{c_3}{2} \langle u^2\rangle
                         + \ldots \right]\psi_N~, \label{s122} \\
{\cal L}_{\pi N}^{(3)}&=&\bar{\psi}_N\left\{
                         B_{23} \, \Delta_0\, \langle\chi_+\rangle
                         +B_{32}\,\Delta_0^3
                         + ...\right\}\psi_N ~,\\ \label{s124}
{\cal L}_{\pi N}^{(4)}&=&\bar{\psi}_N\left\{ 
                         e_{38} \, \langle \chi_+ \rangle^2  
                         +e_{115} \, \frac{1}{4} \langle  \chi_+^2 - \chi_-^2 
                         \rangle 
                         -e_{116} \, \frac{1}{4} \biggl(  \langle  \chi_-^2 
                         \rangle -\langle  \chi_- \rangle^2 +  \langle  
                         \chi_+^2 \rangle - \langle  \chi_+ \rangle^2\biggr)
                         \right. \nonumber \\
                      & & \qquad\qquad\qquad\qquad\qquad\qquad\left.
                         +E_1\, \Delta_0^4
                         +E_2\, \Delta_0^2\,\langle\chi_+\rangle
                         + \ldots \right\}\psi_N ~.
\end{eqnarray}
Here, $\psi_N$ denotes the nucleon spinor, 
$g_A$ is the axial-vector coupling constant (in the chiral limit), $m_0$
the chiral limit value of the nucleon mass, $\chi_+$ contains the external
scalar fields which contains the quark mass matrix and $\langle \ldots\rangle$ 
denotes the trace in flavor space. The $c_i$ are
low-energy constants (LECs). Their values can e.g. be determined in the
analysis of pion-nucleon scattering in the delta-full EFT \cite{FMdel}. It is
known that in particular $c_2$ and $c_3$ are largely saturated by
$\Delta$-exchange \cite{BKMlec}, thus the remaining contributions in an EFT
with explicit spin-3/2 fields are expected to be very small. 
The counterterms $B_{23}, \, B_{32}, \, E_1, \, E_2$ 
in ${\mathcal L}_{\pi N}^{(3,4)}$ appear naturally in SSE ($\Delta_0 \sim
{\cal O}(\epsilon)$) and are required for the 
renormalization of the loop graphs at ${\cal O}(\epsilon^{3,4}$). 
Furthermore, their finite parts are chosen in such a way that in the 
limit $m_\pi /\Delta_0 \rightarrow 0$ the ${\cal O}(\epsilon^n)$ SSE result
recovers the ${\cal O}(p^n)$ result of heavy baryon chiral perturbation
theory (HBCHPT). The  fourth-order LECs $e_i$
contribute to the nucleon mass in the combination $e_1 = 16e_{38}+2e_{115}
+ e_{116}/2$ (for details, see \cite{FMMS,BHMcut}). The relevant terms from
the covariant $\pi N \Delta$ Lagrangian read  \cite{Hemmert}
\begin{eqnarray}
{\cal L}_{\pi N\Delta}^{(1)}&=&c_A\bar{\psi}^i_\alpha O^{\alpha\beta}_{A} w_\beta^i \psi_N
                               + {\rm h.c.}~,   \label{s12321} \\
{\cal L}_{\pi N\Delta}^{(2)}&=&\bar{\psi}^i_\alpha  O^{\alpha\mu}_{A} \left\{
                               b_3\,i\,w_{\mu\nu}^i\,\gamma^\nu
                               +\frac{b_6}{m_0}\,i\,w_{\mu\nu}^i\,i\,D^\nu
                               + \ldots \right\} \psi_N+ {\rm h.c.}~,
\end{eqnarray}
with
$w_\mu^i = \langle\tau^i u_\mu\rangle/2$ and 
$w_{\mu\nu}^i =  \langle\tau^i \left[D_\mu,u_\nu\right]\rangle/2$,
where $\tau^i,\,i=1,2,3$ denote the Pauli matrices in isospin space. The tensor 
$O_{\alpha\beta}^{A}=g_{\alpha\beta}+\frac{2\,A}{d}\gamma_\alpha\gamma_\beta$ 
takes care of the ``point-invariance" of the spin-3/2 
theory \cite{HHK}. 
Here, ${\psi}_i^\mu$ denotes the spin-3/2 field in 
Rarita-Schwinger notation with an additional isospin-index $i$ and the constraint 
$\tau^{i} \,{\psi}_i^\mu=0$ \cite{HHK}. The LEC  
$c_A$ represents the leading axial-vector  $N \Delta$ coupling constant 
(frequently called $g_{\pi N \Delta}$ in the literature). At second order,
we have two relevant couplings, parameterized by the LECs $b_3$ and $b_6$. We note 
that in the non-relativistic theory the leading contribution of $b_6$ can be 
absorbed in the coupling 
$b_3$ to ${\cal O}(\varepsilon^2)$ \cite{HHK}. In the non-relativistic SSE $b_6$
therefore only appears as an independent coupling
 at ${\cal O}(\varepsilon^3)$ . Finally, we list the 
terms of the Lagrangian that describe the delta propagation and its
coupling to pions and external sources. The pertinent covariant 
structures up to ${\cal O}(\varepsilon^2)$ have been constructed 
in \cite{Hemmert},  here we add further terms needed up  ${\cal
  O}(\varepsilon^4)$ in close analogy to the nucleon case  (see also \cite{FMdel}).
 Specifically, we employ
\begin{eqnarray}
{\cal L}_{\pi\Delta}^{(1)}&=& \!\!\!\!-\bar{\psi}_\alpha^i O^{\alpha\mu}_{A} \left\{
                                               \left[i \not{\!\!D}^{ij}
        -m_{\Delta}^0\xi^{ij}_{3/2}+\frac{g_1}{2}\!\!\not{\!\!u}^{ij} \gamma_5\right] g_{\mu\nu}
        -\frac{\gamma_\mu\gamma_\lambda}{4} 
         \left(i \not{\!\!D}^{ij}-m_{\Delta}^0\xi^{ij}_{3/2}\right)\gamma^\lambda
                       \gamma_\nu \right\} O^{\nu\beta}_{A} \psi_\beta^j,
                                               \label{s321}\\ 
%                    \nonumber\\ && \\
{\cal L}_{\pi\Delta}^{(2)}&=&  -\,\bar{\psi}_\alpha^i O^{\alpha\mu}_{A} 
                   \left\{\left[a_1 \langle \chi_+\rangle\delta^{ij}
                   -\frac{a_2}{4 m_0^2}\left\{\langle u_\alpha u_\beta \rangle 
                          D^\alpha_{ik} D^\beta_{kj}+ {\rm
                   h.c.}\right\}\right.\right. \nonumber \\
                 && \left.\left.  \quad\qquad\qquad\qquad\qquad\qquad 
                         +\frac{a_3}{2} \langle u^2\rangle\delta^{ij}+ \ldots \right]g_{\mu\nu}
                                              + \ldots \right\} O^{\nu\beta}_{A} \psi_\beta^j, \\
{\cal L}_{\pi\Delta }^{(3)}&=&-\, \bar{\psi}_\alpha^i O^{\alpha\mu}_{A} \left\{
                         B_1^\Delta \, \Delta_0\,\langle \chi_+\rangle
                         +B_0^\Delta\,\Delta_0^3
                         + \ldots \right\}g_{\mu \nu} \delta^{ij} O^{\nu\beta}_{A} \psi_\beta^j \\
{\cal L}_{\pi \Delta}^{(4)}&=&-\, \bar{\psi}_\alpha^i O^{\alpha\mu}_{A} \left\{ 
                         e_{38}^\Delta \, \langle \chi_+ \rangle^2  
                         +e_{115}^\Delta \, \frac{1}{4} \langle  \chi_+^2 
                          - \chi_-^2 
                         \rangle 
                         -e_{116}^\Delta \, \frac{1}{4} \biggl(  \langle  
                          \chi_-^2 
                         \rangle -\langle  \chi_- \rangle^2 +  \langle  
                         \chi_+^2 \rangle - \langle  \chi_+ \rangle^2\biggr)
                         \right. \nonumber \\
                      & &\left. \quad\qquad\qquad\qquad\qquad
                         +E_1^\Delta\, \Delta_0^4
                         +E_2^\Delta\, \Delta_0^2\, \langle \chi_+\rangle
                         + \ldots \right\}g_{\mu \nu} \delta^{ij} O^{\nu\beta}_{A} \psi_\beta^j
\end{eqnarray}
with
$D_\mu^{ij} = D_\mu\delta^{ij}-i\epsilon^{ijk}\, \langle\tau^k D_\mu\rangle$,
$u_\mu^{ij} = u_\mu\delta^{ij}$ and
$\xi^{ij}_{3/2}$ denotes the isospin 3/2 projector, $\xi^{ij}_{3/2} =
\delta^{ij}-\tau^i\tau^j/3$.
Further, $g_1$ denotes the axial $\Delta\Delta$ coupling constant (in the chiral limit), 
whereas $m_\Delta^0$  refers to the $\Delta$(1232) mass (in the chiral limit). The LECs
$a_i$ correspond\footnote{We note that while only three of the $c_i$ LECs are 
contributing to the mass functions considered here, 
there are actually 9 different LECs $a_i$ that should be taken into account. 
However, it can be shown that there are only three different linear
combinations of them contributing to the masses. We therefore take
$a_1,\,a_2,\,a_3$ as representatives of these three 
independent structures.}
to the $c_i$ in Eq.~(\ref{s122}) and so on. The $B_i^\Delta$ and $E_i^\Delta$
are taken such that the values of $m_0^\Delta$ and $a_1$ {\em in the chiral
  limit} are not affected by loop corrections. 

\medskip

%%%%%%%%%%%%%%%%%%%%%%%%%%%%%%%%%%%%%%%%%%%%%%%%%%%%%%%%%%%%%%%%%%%%%%%%%%%%%%%%%%%%%%%%
\noindent {\bf 3.} Choosing $A=-1$ the propagator for a spin-3/2--isospin-3/2
particle in the Rarita-Schwinger formalism has 
the general form\footnote{The small scale expansion constitutes one possible form 
of an effective low energy theory of QCD for low lying baryon resonances. 
Its range of validity and applicability therefore clearly lies in the low energy 
domain with momenta below the chiral symmetry breaking scale $\Lambda_\chi$. 
All physics connected with higher energies, e.g. the field theoretical
deficiencies 
of the Rarita-Schwinger approach discussed 
in the literature \cite{Johnson},
is therefore mandated to be accounted for  only via the counterterms/local 
operators of the theory. 
Gauge invariance  can be shown to hold perturbatively for all forms of 
the propagators addressed in this work, see Ref.\cite{resonanceSSE}. For alternative approaches
 to spin-3/2 particles in ChEFT see \cite{Dashen,Vladimir}.}
\begin{eqnarray}
S_{\mu\nu}^{ij}(p)&=&-i\,\frac{\not{p}+m_{\Delta}^0}{p^2-(m_\Delta^0)^2}\,
\left[g_{\mu\nu} - { 1 \over d-1} \, \gamma_\mu \gamma_\nu 
-{ (d-2)\, p_\mu p_\nu \over (d-1)\, (m_\Delta^0)^2}
+ { p_\mu \gamma_\nu - p_\nu \gamma_\mu \over (d-1) \,
  m_\Delta^0}\right]\,\xi^{ij}_{3/2}\; . 
\label{RS}
\end{eqnarray}
The Dirac-tensor $S_{\mu\nu}$ can be written as a linear superposition of 
spin-3/2 and spin-1/2 projection operators $P_{\mu\nu}^{3/2},\,P_{\mu\nu}^{1/2}$:
\begin{eqnarray}
-i\,S_{\mu\nu}^\Delta (p) 
&=& -{\not{p} + m_\Delta^0 \over p^2 - (m_\Delta^0)^2}\, P_{\mu\nu}^{3/2}
- {1 \over \sqrt{d-1} m_\Delta^0} \Biggl( \bigl( P_{12}^{1/2}\bigr)_{\mu\nu} + 
\bigl( P_{21}^{1/2}\bigr)_{\mu\nu} \Biggr) \nonumber \\ &&\quad + 
{d-2\over (d-1)\, (m_\Delta^0)^2} \, (\not{p} + m_\Delta^0 )
\, \bigl( P_{22}^{1/2}\bigr)_{\mu\nu}~. \label{general}
\end{eqnarray}
For details regarding the properties of the projectors we refer to \cite{SSEIR}. 
As can be clearly seen from Eq.(\ref{general}) only the spin-3/2 components are 
associated with a propagation, whereas the (spurious) spin-1/2 components correspond 
to local contact operators. Given that the chiral effective field theory for a 
coupled nucleon-delta system contains the most general set of local contact operators 
allowed by chiral symmetry, it was argued in \cite{SSEIR} that in chiral EFT one only 
needs to take into account the propagation of the spin-3/2 degrees of freedom. 
The effects of the spurious/off-shell spin-1/2 components are already completely 
accounted for by the counterterms of the theory.  
In \cite{SSEIR} we have therefore proposed to utilize the propagator form A 
\begin{eqnarray}
G_{\mu\nu}^{ij}|_{\rm IR}^A(p)&=&-i\,\frac{\not{p}+m_{\Delta}^0}{p^2-(m_\Delta^0)^2}\,
P_{\mu\nu}^{3/2}\,\xi^{ij}_{3/2}\; , \label{formA}
\end{eqnarray}
with the spin-3/2 projector
\begin{eqnarray}
P_{\mu\nu}^{3/2}&=&g_{\mu\nu}-\frac{1}{d-1}\,\gamma_\mu\gamma_\nu
-\frac{1}{(d-1)\,p^2}\left(\not{p}\gamma_\mu p_\nu+p_\mu\gamma_\nu\not{p}\right)
-\frac{(d-4)}{(d-1)}\,\frac{p_\mu p_\nu}{p^2}~. \label{s3/2}
\end{eqnarray}
In the meantime it was realized in practical applications that it is advantageous to 
use a different form for the propagator than the one given in Eq.(\ref{formA}) (see
also the related work in \cite{MZdel}). 
In Ref.\cite{SSEIR} it was shown that due to the extra $1/p^2$ structure in 
Eq.(\ref{s3/2}) the standard rules of infrared regularization had to be modified 
for a consistent treatment of the low-energy modes, making loop-calculations in this 
covariant field theory  more tedious. Here we propose another separation of the 
general Dirac-tensor:
\begin{eqnarray}
S_{\mu\nu}^\Delta&=&-i\,\frac{\not{p}+m_{\Delta}^0}{p^2-(m_\Delta^0)^2}\,
\frac{p^2}{(m_\Delta^0)^2}\,P_{\mu\nu}^{3/2}+i\,R_{\mu\nu}
\end{eqnarray} 
Once more it can be shown that the remainder 
\begin{eqnarray}
R_{\mu\nu}&=&\frac{\not{p}+m_\Delta^0}{(m_\Delta^0)^2}
\left[g_{\mu\nu}-\frac{1}{d-1}\,\gamma_\mu\gamma_\nu\right]
+\frac{1}{(d-1)\,(m_\Delta^0)^2}\left[p_\mu\gamma_\nu-\gamma_\mu p_\nu\right]
\end{eqnarray}
is not connected with the spin-3/2 propagation, that it is local and that it cannot generate 
any contributions which are not already contained in the counterterms of 
the effective field theory. As an alternative to Eq.(\ref{formA}) 
we therefore propose\footnote{We note that this form of the spin-3/2 
propagator is also employed when the interactions of the spin-3/2 field
are constructed according to the rules of gauge symmetry for a massless
spin-3/2 field, e.g. see \cite{Vladimir}.} the propagator form
\begin{eqnarray}
G_{\mu\nu}^{ij}|_{\rm IR}^B(p)&=&-i\,\frac{\not{p}+m_{\Delta}^0}{p^2-(
                             m_\Delta^0)^2}\,\frac{p^2}{(m_\Delta^0)^2}\,
                             P_{\mu\nu}^{3/2}\,\xi^{ij}_{3/2}\; . \label{formB}
\end{eqnarray}
We note that in form B there is no more $1/p^2$ structure, making the evaluation 
of loop diagrams containing spin-3/2 degrees of freedom in the IR-scheme a lot simpler. 
However, we emphasize that in a low-energy (chiral) effective field theory the 
finite/renormalized results obtained via Eq.(\ref{formA}) 
or Eq.(\ref{formB}) are identical\footnote{Obviously the high-energy behavior 
between the two propagator expressions is different, leading
 to different numerical values for the associated counterterms to yield the
 same total result.}---it is merely a matter of practicability which method 
is chosen for a particular observable. In what follows, we will work with
form~B. We note explicitly that one could also work 
with the full Rarita-Schwinger propagator of Eq.(\ref{RS}), however, form~A or 
form~B have the advantage that the results can be directly mapped
to the known couplings of the heavy-baryon formulation of SSE where also only
the spin-3/2 components are part of the propagator \cite{JMdel,HHK}. 
Finally we remark that for low energies the width of $\Delta(1232)$ can be treated 
perturbatively in the covariant SSE, analogous to the situation in the
non-relativistic theory. Throughout this work we will therefore not address
 this issue, applications of the covariant SSE in the resonance region will be
 discussed in a forthcoming paper \cite{resonanceSSE}.

\medskip

\begin{figure}[tb]
\parbox{.49\textwidth}{\epsfig{file= 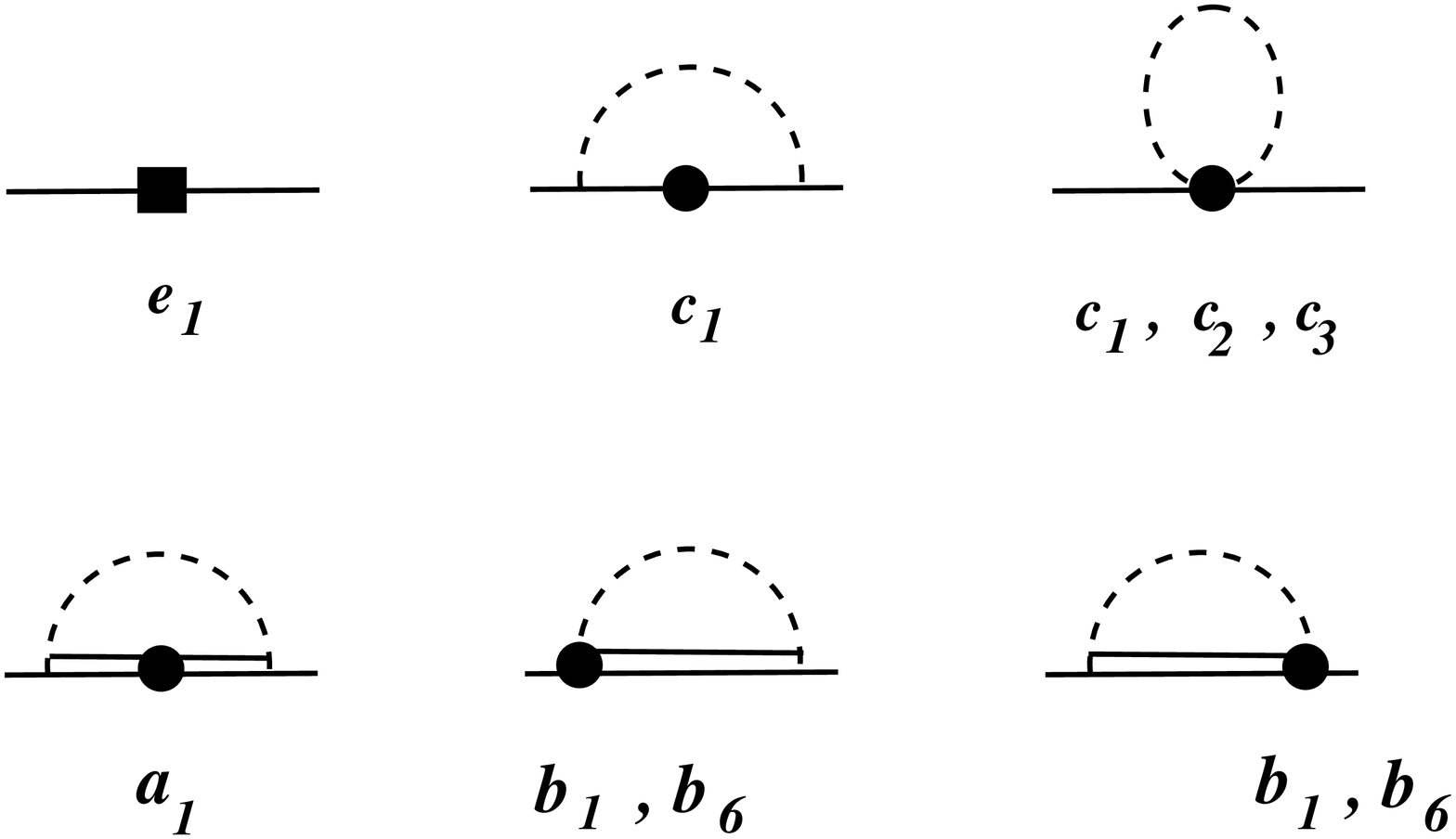,width=.44\textwidth,silent=,clip=}}
\hfill
\parbox{.49\textwidth}{\epsfig{file= 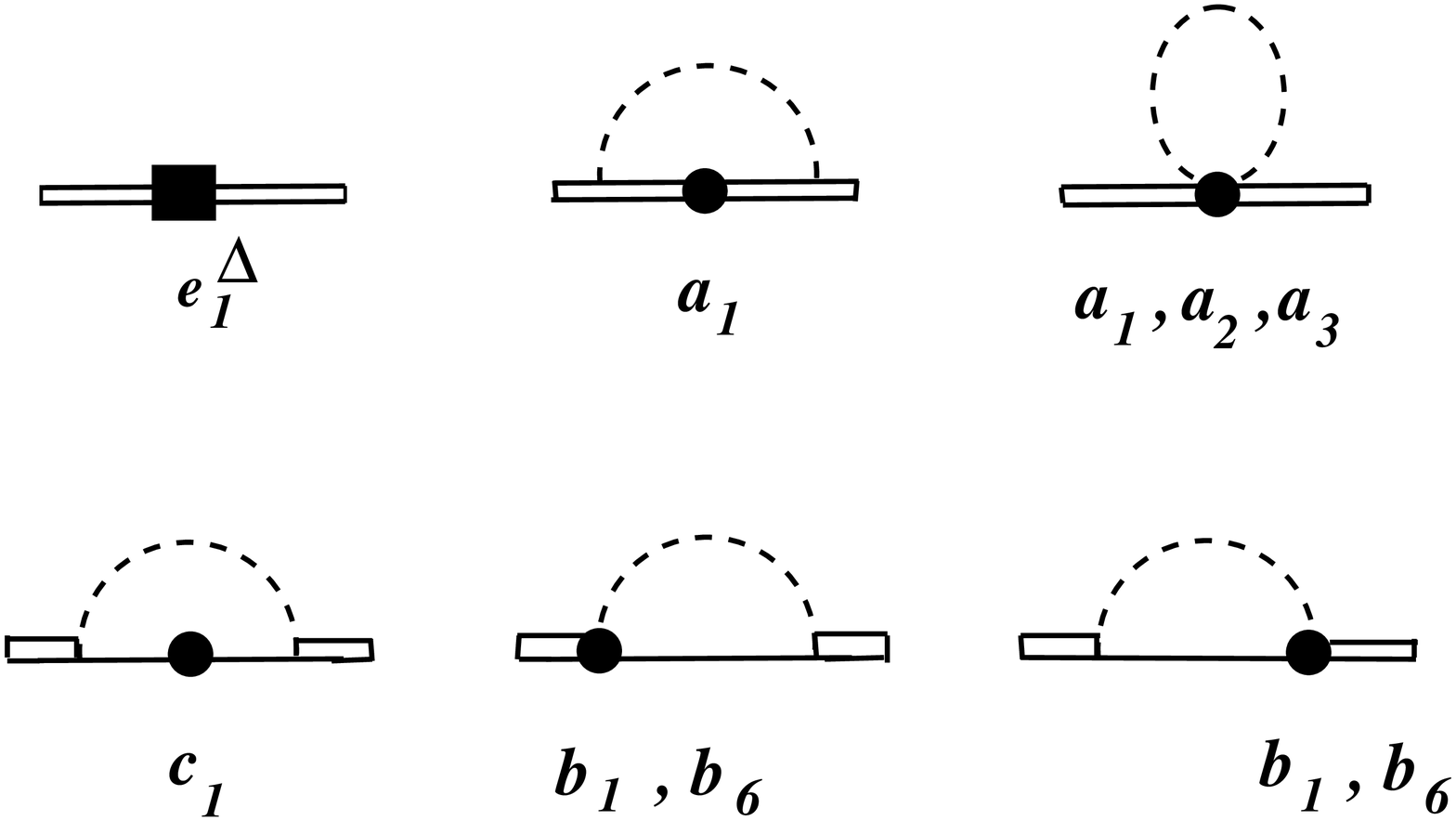,width=.44\textwidth,silent=,clip=}}
\vspace{0.2cm}
\begin{center}
\caption{Fourth order contributions to the nucleon (left panel) and the
delta mass (right panel). Solid, double and dashed lines denote nucleons,
deltas and pions, in order. The corresponding dimension two $(a_i, b_i, c_i)$
and four ($e_i$) LECs are also given at the pertinent vertices.
\label{fig:dia}}
\vspace{-0.3cm}
\end{center}
\end{figure}

\noindent {\bf 4.} Armed with the effective Lagrangian, we are now in the
position to calculate the nucleon and the delta self-energy to fourth order
in the covariant SSE. The covariant formalism allows in particular to resum
the contributions from kinetic energy insertions. The nucleon mass can be written as
\begin{equation}\label{mn}
m=m_0-4c_1M_\pi^2-4M_\pi^2\Delta_0B_{23} -B_{32}\Delta_0^3-4 e_1(\lambda)  M_\pi^4 
 -E_1 \Delta_0^4 -4 M_\pi^2 \Delta_0^2 E_2  + m^{{\rm N-loop}}
+  m^{{\rm \Delta-loop}}~, 
\end{equation}
with  the corresponding fourth-order tree and one-loop graphs shown in
Fig.~\ref{fig:dia}. The contributions from the nucleon and delta loop graphs,
which start at ${\mathcal O}(\ve^3)$, take the form
\begin{eqnarray}             
m^{{\rm N-loop}} &=&-  \frac{3g_A^2}{2F_\pi^2}m_0 M_\pi^2I_{11}^{(0,0)}(m_0^2,m_0^2)
-\frac{3M_\pi^2}{ F_\pi^2} \left(-2c_1+\frac{c_2}{d}+c_3\right) \Delta_\pi~, \\
%\end{eqnarray}
%\begin{eqnarray}
m^{{\rm \Delta-loop}}&=&
\frac{2c_A^2\,(d-2)}{F_\pi^2}\frac{m_0^2}{(m_\Delta^0)^2}\,
\left\{\left(m+m_\Delta^0\right)
I_{11}^{(2,0)}(m_0^2,(m_\Delta^0)^2)
          -m_0 I_{11}^{(3,1)}(m_0^2,(m_\Delta^0)^2)\right\}
\nonumber \\
& -& b_3 \frac{c_A}{2 F_\pi^2 (m_\Delta^0)^2} \frac{d-2}{d-1} \biggl\{
\biggl[-(m_0^2-
(m_\Delta^0)^2)^2+M_\pi^2\biggl(4 \frac{d-1}{d} m_0^2 -4 \frac{m_0  m_\Delta^0}{d}
+M_\pi^2\biggl)\biggr]\Delta_\pi \nonumber \\
& -& \bigl[(m_0^2-(m_\Delta^0)^2)^2+M_\pi^2(M_\pi^2-2(m_0^2+
(m_\Delta^0)^2))\bigr]
\nonumber \\
& \times &\biggl[(m_0^2-(m_\Delta^0)^2-M_\pi^2)
I_{11}^{(0,0)}(m_0^2,(m_\Delta^0)^2)+2m_0 m_\Delta^0
I_{11}^{(1,1)}(m_0^2,(m_\Delta^0)^2) \biggr] \biggr\}
\nonumber \\
&- &\frac {b_6}{m_0}   \frac{c_A}{2 F_\pi^2 (m_\Delta^0)^2}  \biggl\{
\biggl[ -(m_0^2-
(m_\Delta^0)^2)^2+M_\pi^2\biggl(2 \frac{d-2}{d} m_0^2 +2 (m_\Delta^0)^2\biggr)-M_\pi^2]
(m_0+
 m_\Delta^0) \nonumber \\
& -&  \frac{2}{d}M_\pi^2 m_0(-2M_\pi^2-m_0^2+  (m_\Delta^0)^2) \biggr]
\Delta_\pi- \bigl[(m_0^2-(m_\Delta^0)^2)^2+M_\pi^2(M_\pi^2-2(m_0^2+
(m_\Delta^0)^2)  \nonumber \\
& \times & (m_0^2-(m_\Delta^0)^2+M_\pi^2)((m_0 +m_\Delta^0)
I_{11}^{(0,0)}(m_0^2,(m_\Delta^0)^2) -m_0
I_{11}^{(1,1)}(m_0^2,(m_\Delta^0)^2) \bigr] \biggr\}~.
\end{eqnarray}
Here, $F_\pi$ is the pion decay constant and $\lambda$ the scale of
dimensional regularization. Throughout, we set $\lambda = 1.232$~GeV.
\noindent
The loop functions which appear here and for the delta mass discussed below
are defined as:
\begin{eqnarray}
& &\frac{1}{i} \int d^4 k \{1,k_\mu,k_\mu k_\nu, k_\mu k_\nu k_\rho \} 
\frac{1}{(M_\pi^2-k^2)(m^2-(p-k)^2)} = %\nonumber \\
%& &  
\biggl\{ H_{11}^{(0,0)}(p^2,m^2) \ , \nonumber \\
&& p_\mu H_{11}^{(1,1)}(p^2,m^2),
g_{\mu \nu} H_{11}^{(2,0)}(p^2,m^2) + \ldots, 
(p_\rho g_{\mu \nu} + p_\mu g_{\rho \nu} %\nonumber\\
%&+&
+ p_\nu g_{\mu \rho} H_{11}^{(3,1)}(p^2,m^2) +\ldots \biggr\}~.
\end{eqnarray}
Their irregular parts are given by:
\begin{equation}
I_{11}^{(0,0)}(p^2,m^2)=
  -\frac{p^2-m^2+M_\pi^2}{2 p^2} \frac {1}{16 \pi^2}
\left({\ln}\frac{M_\pi^2}{p^2}-1\right) - \bar J_0
\end{equation}
with
\begin{equation}
%\hskip -1truecm
\bar J_0 = \frac{1}{32 \pi^2 p^2} \
 \sqrt{\lambda(p^2,m^2,M_\pi^2)} \ {\ln}\frac {p^2-m^2+
M_\pi^2+ \sqrt{\lambda(p^2,m^2,M_\pi^2)}}{p^2-m^2+
M_\pi^2- \sqrt{\lambda(p^2,m^2,M_\pi^2)}}
+i  \frac {\sqrt{\lambda(p^2,m^2,M_\pi^2)}}{16 \pi p^2} ~,
\end{equation}
if $\lambda(p^2,m^2,M_\pi^2)>0$. Otherwise,  we have
\begin{equation}
\bar J_0 = \frac{1}{16 \pi^2 p^2} \
\sqrt{-\lambda(p^2,m^2,M_\pi^2)} \ \ {\arccos} \left(- \frac{p^2-m^2+M_\pi^2}
{2 M_\pi \sqrt{p^2}}\right)~.
\end{equation}
The other relevant loop functions can be expressed as
\begin{eqnarray}
I_{11}^{(1,1)}(p^2,m^2) &=& \frac{1}{2 p^2}\biggl( (p^2-m^2+M_\pi^2) I_{11}^{(0,0)}
+\frac{1}{2} \Delta_\pi \biggr)~, 
\\
I_{11}^{(2,0)}(p^2,m^2) &=& \frac{1}{4(d-1) p^2}\biggl(\lambda(p^2,m^2,M_\pi^2)
  I_{11}^{(0,0)} -(p^2-m^2+M_\pi^2) \Delta_\pi \biggr)~, 
\\
I_{11}^{(3,1)}(p^2,m^2) &=& \frac{1}{2 p^2}\biggl((p^2-m^2+M_\pi^2)
  I_{11}^{(2,0)} +\frac{M_\pi^2}{d} \Delta_\pi \biggr) ~.
\end{eqnarray}
We also employ the standard definition of $\Delta_\pi$,
\begin{equation}
\Delta_\pi = \frac{1}{i} \int \frac {d^d k}{(2 \pi)^d} \frac{1}{M_\pi^2-k^2}~.
\end{equation}
Using these,
the $\Delta$-loop contribution can be written as an infinite series in $M_\pi$
(modulo chiral $\log$s):
\begin{equation} 
m_N^{\rm \Delta-loop}= l_0 +l_1\, M_\pi^2 + l_2 \, M_\pi^4 + \ldots~,
\end{equation}
with $l_0, l_1$ being infinite series in $\Delta_0$ starting at 
${\mathcal O}(\ve^3)$ and ${\mathcal O}(\ve)$, respectively. Note that the 
decoupling theorem is automatically fulfilled here since there is
no nonanalytic contribution $\sim M_\pi^3$.  Now as a minimal way of determining
the LECs $B_{23}$, $B_{32}$, $E_1$ and $E_2$ one can choose
$B_{23}+ \Delta_0^2 E_2 =l_1/4$, and $B_{32}+\Delta_0 E_1 = l_0/4$,
which means that the $\Delta$ contribution will first occur at order $M_\pi^4$.

\smallskip\noindent
The chiral expansion of the delta mass can be written in a way similar
to the case of the nucleon Eq.~(\ref{mn}) (the corresponding tree and
loop graphs that contribute at ${\cal O}(\ve^4)$ are collected in
Fig.~\ref{fig:dia}),
\begin{equation}\label{mdelta}
m_\Delta=m_\Delta^0-4a_1M_\pi^2-4M_\pi^2\Delta_0B_{1}^\Delta
    -B_{0}^\Delta\Delta_0^3 -4 e_1^\Delta(\lambda) M_\pi^4  
-E_1^\Delta \Delta_0^4 -4 M_\pi^2 \Delta_0^2 E_2^\Delta
+m_\Delta^{{\rm N-loop}}+m_\Delta^{{\rm \Delta-loop}}~,
\end{equation}
with
\begin{eqnarray}
m_\Delta^{{\rm N-loop}}&=&
\frac{c_A^2}{F_\pi^2}\,\left\{\left(m_\Delta^0+m_0\right)I_{11}^{(2,0)}
((m_\Delta^0)^2,m_0^2)-m_\Delta^0 I_{11}^{(3,1)}((m_\Delta^0)^2,m_0^2)\,\right\} 
\nonumber  \\
& + & 
b_3 \frac{c_A}{F_\pi^2}\frac{M_\pi^2}{d} \biggl\{ \Delta_\pi+[(m_\Delta^0)^2-
m_0^2-M_\pi^2]I_{11}^{(0,0)}((m_\Delta^0)^2,m_0^2)+2 m_0 m_\Delta^0 
 I_{11}^{(1,1)}((m_\Delta^0)^2,m_0^2) \biggr\}
\nonumber \\
& + &  \frac{b_6}{m_0} \frac{c_A}{F_\pi^2}\frac{M_\pi^2}{d} \biggl\{ 
\Delta_\pi (m_0+m_\Delta^0)+
[(m_\Delta^0)^2 - m_0^2+M_\pi^2]
\nonumber \\
& \times &\biggl[(m_0+m_\Delta^0)  I_{11}^{(0,0)}((m_\Delta^0)^2,m_0^2)- m_\Delta^0  
I_{11}^{(1,1)}((m_\Delta^0)^2,m_0^2)\biggr] \biggr\}
\end{eqnarray}
and
\begin{eqnarray}
m_\Delta^{{\rm \Delta-loop}}
& = & -\frac{5 h_A^2}{6 (m_\Delta^0)^2 F_\pi^2}\,
\left\{ m_\Delta^0
\biggl[ M_\pi^2 \Delta_\pi \biggl(1-\frac{4}{d (d-1)}\biggr)-M_\pi^2 (m_\Delta^0)^2 
I_{11}^{(0,0)}((m_\Delta^0)^2,(m_\Delta^0)^2) \right.\nonumber \\
&+ &\frac{1}{d-1}(4 (m_\Delta^0)^2 +(d-4) M_\pi^2)  I_{11}^{(2,0)}
((m_\Delta^0)^2,(m_\Delta^0)^2)
\biggr] + m_\Delta^0\biggl[\frac{4}{d-1} (m_\Delta^0)^2 I_{11}^{(3,1)}
((m_\Delta^0)^2,(m_\Delta^0)^2)
\bigr. \nonumber \\
& -& \left. \bigl. \frac {M_\pi^2}
{d} \Delta_\pi \biggr]
-\frac{1}{d (d-1)} M_\pi^2 \Delta_\pi \right\} 
-\frac{3 M_\pi^2}{ F_\pi^2}\left(-2a_1+\frac{A}{d}+B\right) \Delta_\pi~,
\end{eqnarray}
with $A=a_2+\dots$ and $B=a_3+\ldots$.
It can easily be verified that $m_\Delta^{{\rm \Delta-loop}}$ starts as $M_\pi^3$ whereas, 
as in the case of the nucleon mass, $m_\Delta^{\rm N-loop}$ has a constant 
and a piece $\sim M_\pi^2$:
\begin{equation}
 m_\Delta^{\rm N-loop}= l_0+l_1M_\pi^2 + l_2 M_\pi^4 + \cdots
\end{equation}
As before,  we will choose $B_0^\Delta$, $B_1^\Delta$, $E_1^\Delta$
$E_2^\Delta$ in  such a way that $ m_\Delta^{\rm N-loop}$
starts to contribute only at $M_\pi^4$, that is
$B_0^\Delta+\Delta_0^2 E_2^\Delta = l_1/4$ 
and $B_1^\Delta + \Delta_0 E_2^\Delta = l_0/4$. Note that we did not take into account the
difference between the chiral limit value of the pion mass and its physical
value \cite{Fuhrer}. This effect is known to be  small, see e.g. the
three-flavor calculation \cite{FMS}.

\medskip

\begin{figure}[tb]
\centerline{
\epsfig{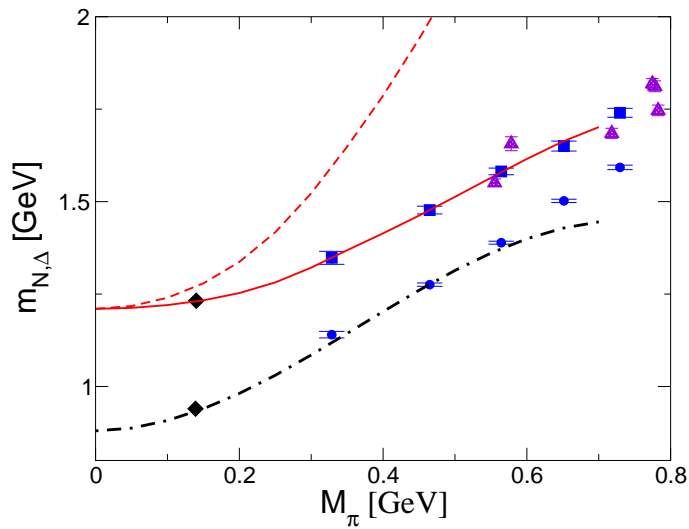}
}
\vspace{0.2cm}
\begin{center}
\caption{The nucleon mass (dot-dashed line) and the (real part of the) delta mass (solid line)
as a function of the pion mass. The filled diamonds denote their physical
values at  the physical pion mass. The dashed line is the chiral extrapolation
for the $\Delta$ based on SU(6) as explained in the text. The filled squares
and circles are the MILC data \protect\cite{MILC}. The filled triangles are
the recent data from QCDSF \protect\cite{QCDSF}.
\label{fig:extra}}
\vspace{-0.5cm}
\end{center}
\end{figure}

\noindent {\bf 5.} We are now in the position to analyze the nucleon
and delta mass formulae given in Eqs.~(\ref{mn},\ref{mdelta}). They
contain a certain number of LECs, some of which are (not very accurately)
known from the study of pion-nucleon scattering in the heavy baryon SSE
\cite{FMdel}. Here our aim is modest\footnote{Ultimately global, simultaneous
  fits of several nucleon and delta observables calculated within
covariant SSE to next-to-leading one-loop order need to be undertaken to 
obtain reliable information on LECs from lattice QCD.}: 
In addition to the known values at the physical point 
we take the data from MILC \cite{MILC}
for the nucleon and the delta as function of the pion mass and try to
describe these with LECs of natural size. Such a description is indeed
possible, as shown in Fig.~\ref{fig:extra}. So we do not intend detailed
least-square fits here but rather try to find out whether the existing data shown in the
this figure can be consistently described by our mass formulas with LECs of
natural size. We stress again that a more refined analysis of
e.g. pion-nucleon scattering in the covariant SSE is mandatory to put
stringent constraints on certain combinations of the LECs (see also the
extensive discussion in  Refs.\cite{BHMcut,HPW} on this issue).
\noindent
We have 11 (combinations of) parameters to determine, these are 
$c_1, c_2/4+c_3, a_1 ,e_1, e_1^\Delta,  c_A, h_A, b_3, b_6,
A/4+B$ and  $\Delta_0$. \underline{(i) constrained parameters:} 
$c_A$ and $\Delta_0$ are fixed from the mass and
width of the delta. We note that the $N\Delta$ mass splitting in the chiral limit, 
$\Delta_0  = 0.33\,$GeV, indicates a slightly {\it larger} $N\Delta$ mass
splitting in the chiral limit than at the physical point. A similar
observation was made in the case of quenched QCD, see \cite{James} (for 
discussion about the difference between the mass splitting in the quenched
approximation and full QCD, see also \cite{young}). 
The LECs from the pion-nucleon Lagrangian, Eqs.~(\ref{s122},\ref{s124}) 
can be estimated from pion-nucleon scattering in the presence of the delta.
We use $c_1=-0.8\,$GeV$^{-1}$ (which is within the
uncertainty of the values determined in e.g. \cite{BuM}) and 
$e_1 = c_2 = 0$, $c_3=0.5\,$GeV$^{-1}$. The small values of $c_{2,3}$ are
consistent with resonance saturation studies of \cite{BKMlec} and the
fits in \cite{FMdel}. 
%The fourth-order LEC $e_1$ induces the largest uncertainty,
%as discussed in detail in \cite{BHMcut} (even a small value of $e_1$ leads to
%a sizeable contribution at larger pion masses). 
(ii) \underline{Fit parameters:} The remaining parameters are fit to the
masses. We find for
the two axial dimension two $N \Delta$ LECs the values
$b_3 = 0.75$ GeV$^{-1}$ and $b_6 = -0.75$. The 
axial $\Delta \Delta$ coupling is found to be $h_A=2$, which is not far from the
SU(6) or large-$N_C$ value $h_A = 9 g_A/5 = 2.28$. Furthermore we get 
$e_1^\Delta =-1\,$GeV$^{-3}$, $a_1 = -0.3\,$GeV$^{-1}$, 
$A=0$ and  $B=0.5\,$GeV$^{-1}$.
These are all natural values. It is interesting to note that $a_1$ is markedly
smaller than $c_1$, although both couplings should be equal in the SU(6) limit. We refrain
here from a detailed study of the theoretical errors that will be given in
a forthcoming publication. Still, it is interesting to study the strict SU(6)
limit. In that case, one would have $a_1 = c_1 = -0.8\,$GeV$^{-1}$ and $h_A = 2.28$.
As can be seen from the dashed line in 
Fig.~\ref{fig:extra}, the assumption of strict SU(6) symmetry is clearly at odds
with the MILC data, indicating that $a_1$ and $c_1$ indeed seem to have
different values.  Also shown in Fig.~\ref{fig:extra} are the recent QCDSF data for $m_\Delta$,
which were not used in the fit but are nicely consistent with our
extrapolation function. Note also that the  QCDSF data are based on two-flavor
simulations and are not very different from the MILC data in the region of
overlap. This further supports our assumption on the treatment of the MILC data.
We stress again that the resulting values of the LECs are to
be considered indicative and a more analysis employing also contraints from
other physical processes should follow.

From the small value of the LEC $a_1$ one immediately deduces that the
$\pi\Delta$ sigma term appears to be significantly {\it smaller} than 
its nucleon cousin because
at leading order in the quark mass expansion we have $\sigma_{\pi N} = 
-4c_1 M_\pi^2+...$ and $\sigma_{\pi \Delta} =-4a_1 M_\pi^2+...$.
It is clear that this interesting observation deserves further study. 
Finally we note that the sigma term for 
the nucleon resulting from this ``rough" fit is found as
%35.56~MeV to $\ve^3$ and 
\beq
\sigma_{\pi N} = 48.9~{\rm MeV}~, 
\eeq
to order $\ve^4$. We note that this  value is consistent with the 
classical result of
Ref.~\cite{GLS}, which was confirmed in \cite{BuM} in a HBCHPT analysis of 
pion-nucleon scattering and
in \cite{HPW} in a ChEFT analysis of lattice data in a formalism without 
explicit delta degrees of freedom. It is also in agreement with the recent
CHPT analysis of the three--flavor MILC data, see \cite{FMS}.  For the
$\pi\Delta$ sigma term we get $\sigma_{\pi\Delta} = 20.6\,$MeV. Again,
these results need to be refined and bolstered by more detailed precise
fits to the lattice data including also error and correlation analysis
including also lattice data on other observables - the mass data alone are
not sufficient to precisely pin down all parameters. Such an analysis,
however, goes beyond the scope of this paper.

\medskip

\noindent {\bf 6.} In this letter, we have presented a covariant extension of
the small scale expansion, extending our earlier work \cite{SSEIR}. We have
analyzed the nucleon and the delta mass in view of the lattice data from MILC.
These data seem to indicate  sizeable SU(6) breaking effects
that can be interpreted as a reduced pion cloud contribution in the resonance
field (the delta) as compared to its ground-state (the nucleon).
Our results obtained from fits to full QCD lattice data further indicate that
the $N\Delta$ mass splitting can be larger in the chiral limit than
 at the physical point, analogous to observations in quenched QCD. 
 We have also calculated the pion-nucleon sigma term and found good agreement with earlier
determination from pion-nucleon scattering and the analysis of SU(2) and SU(3)
lattice data for the nucleon. The recent QCDSF data for $m_\Delta$ are also
consistent with our chiral extrapolation. It would be very valuable to have more (2
flavor) lattice results for the delta preferably at small pion masses
to further sharpen these conclusions. 

\vskip 1cm

\noindent{\large {\bf Acknowledgements}}

\smallskip\noindent
TRH acknowledges helpful discussions with G.~Schierholz. 
We also thank G.~Schierholz for providing us with QCDSF data
prior to publication.
TRH is grateful for the hospitality of the LPT at  
Universit\'e Louis Pasteur, Strasbourg, where a large part of 
this work was completed.

%%%%%%%%%%%%%%%%%% REFERENCES %%%%%%%%%%%%%%%%%%%%%%%%%%%%
%%\newpage


\vskip 1cm

\begin{thebibliography}{99}

\frenchspacing
\bibitem{JMdel}
E.~Jenkins and A.~V.~Manohar,
%``Chiral corrections to the baryon axial currents,''
Phys.\ Lett.\ B {\bf 259} (1991) 353.\vs
%%CITATION = PHLTA,B259,353;%%

\bibitem{GZ}
J.~Gasser and A.~Zepeda,
%``Approaching The Chiral Limit In QCD,''
Nucl.\ Phys.\ B {\bf 174} (1980) 445.\vs
%%CITATION = NUPHA,B174,445;%%

\bibitem{HHK}
T.~R.~Hemmert, B.~R.~Holstein and J.~Kambor,
J.\ Phys.\ G {\bf 24} (1998) 1831
[arXiv:hep-ph/9712496].\vs
%%CITATION = HEP-PH 9712496;%%

\bibitem{BFHM}
V.~Bernard, H.~W.~Fearing, T.~R.~Hemmert and U.-G.~Mei{\ss}ner,
%``The form factors of the nucleon at small momentum transfer,''
Nucl.\ Phys.\ A {\bf 635} (1998) 121
[Erratum-ibid.\ A {\bf 642} (1998) 563]
[arXiv:hep-ph/9801297].\vs
%%CITATION = HEP-PH 9801297;%%

\bibitem{BKMrev}
V.~Bernard, N.~Kaiser and U.-G.~Mei{\ss}ner,
Int.\ J.\ Mod.\ Phys.\ E {\bf 4} (1995) 193
[arXiv:hep-ph/9501384].\vs
%%CITATION = HEP-PH 9501384;%%

\bibitem{UGM}
U.-G.~Mei{\ss}ner, in Shifman, M. (ed.): ``At the frontier of particle
physics'', vol. 1, pp. 417-505
(World Scientific, Singapore, 2001) 
%``Chiral QCD: Baryon dynamics,''
[arXiv:hep-ph/0007092].\vs
%%CITATION = HEP-PH 0007092;%%

\bibitem{TRH}
T.~R.~Hemmert, in Proceedings of NSTAR 01, Mainz, Germany; Eds. D. Drechsel and L. Tiator, World Scientific (Singapore) 2002.
%``Chiral effective field theories with explicit spin 3/2 degrees of  
%freedom: A status report,''
[arXiv:nucl-th/0105051].\vs
%%CITATION = NUCL-TH 0105051;%%

\bibitem{BL}
T.~Becher and H.~Leutwyler,
Eur.\ Phys.\ J.\ C  {\bf 9} (1999) 643 
[arXiv:hep-ph/9901384].\vs
%%CITATION = HEP-PH 9901384;%%

\bibitem{FGJS}
T.~Fuchs, J.~Gegelia, G.~Japaridze and S.~Scherer,
%``Renormalization of relativistic baryon chiral perturbation theory and  power counting,''
Phys.\ Rev.\ D {\bf 68} (2003) 056005
[arXiv:hep-ph/0302117].\vs
%%CITATION = HEP-PH 0302117;%%

\bibitem{SSEIR}
V.~Bernard, T.~R.~Hemmert and Ulf-G.~Mei{\ss}ner,
%``Infrared regularization with spin-3/2 fields,''
Phys.\ Lett.\ B {\bf 565} (2003) 137
[arXiv:hep-ph/0303198].\vs
%%CITATION = HEP-PH 0303198;%%

\bibitem{HB}
B.~C.~Tiburzi and A.~Walker-Loud,
%``Strong isospin breaking in the nucleon and Delta masses,''
arXiv:hep-lat/0501018.\vs
%%CITATION = HEP-LAT 0501018;%

\bibitem{aus}
D.~B.~Leinweber, A.~W.~Thomas, K.~Tsushima and S.~V.~Wright,
%``Baryon mass extrapolation,''
Nucl.\ Phys.\ Proc.\ Suppl.\  {\bf 83}, 179 (2000).
[arXiv:hep-lat/9909109].\vs
%%CITATION = HEP-LAT 9909109;%%

\bibitem{James} 
J.~M.~Zanotti, D.~B.~Leinweber, W.~Melnitchouk, A.~G.~Williams and J.~B.~Zhang,
%``Hadron properties with FLIC fermions,''
arXiv:hep-lat/0407039.\vs
%%CITATION = HEP-LAT 0407039;%%

\bibitem{CPPACS}
A.~Ali Khan {\it et al.}  [CP-PACS Collaboration],
%``Light hadron spectroscopy with two flavors of dynamical quarks on the
%lattice,''
Phys.\ Rev.\ D {\bf 65} (2002) 054505 [Erratum-ibid.\ D {\bf 67} (2003) 059901]
[arXiv:hep-lat/0105015].\vs
%%CITATION = HEP-LAT 0105015;%%

\bibitem{JLQCD}
S.~Aoki {\it et al.}  [JLQCD Collaboration],
%``Light hadron spectroscopy with two flavors of O(a)-improved dynamical
%quarks,''
Phys.\ Rev.\ D {\bf 68} (2003) 054502 [arXiv:hep-lat/0212039].\vs
  %%CITATION = HEP-LAT 0212039;%%

\bibitem{QCDSF}
G. Schierholz, private communication; QCDSF collaboration, in preparation.\vs

\bibitem{MN}  
A.~Ali Khan {\it et al.}  [QCDSF-UKQCD Collaboration],
%``The nucleon mass in N(f) = 2 lattice QCD: Finite size effects from  chiral
%perturbation theory,''
Nucl.\ Phys.\ B {\bf 689} (2004) 175 [arXiv:hep-lat/0312030].\vs
  %%CITATION = HEP-LAT 0312030;%% 
 
\bibitem{conferences}
U.-G. Mei{\ss}ner, plenary talk at BARYONS 2004, Paris, France 
[arXiv:hep-ph/0501009], to appear in Nucl. Phys. {\bf A}; T.R. Hemmert, in proceedings of the 
337th WE-Heraeus seminar on Effective Field Theories in
Nuclear, Particle and Atomic Physics (EFT04), J.~Bijnens, U.-G. Mei{\ss}ner and
A.~Wirzba (eds.),  Bad Honnef, Germany [arXiv:hep-ph/0502008].\vs

\bibitem{MILC}
C.~W.~Bernard {\it et al.},
%``The QCD spectrum with three quark flavors,''
Phys.\ Rev.\ D {\bf 64}, 054506 (2001) [arXiv:hep-lat/0104002].\vs
%%CITATION = HEP-LAT 0104002;%%

\bibitem{Karl}
S.~Hashimoto,
%``Recent results from lattice calculations,''
arXiv:hep-ph/0411126.\vs
%%CITATION = HEP-PH 0411126;%%[arXiv:hep-lat/0104002].\vs

\bibitem{BHMcut}
V.~Bernard, T.~R.~Hemmert and U.-G.~Mei{\ss}ner,
%``Cutoff schemes in chiral perturbation theory and the quark mass  expansion
%of the nucleon mass,''
Nucl.\ Phys.\ A {\bf 732} (2004) 149
[arXiv:hep-ph/0307115].\vs
%%CITATION = HEP-PH 0307115;%%

\bibitem{FMS}
M.~Frink, U.-G.~Mei{\ss}ner and I.~Scheller,
%``Baryon masses, chiral extrapolations, and all that,''
arXiv:hep-lat/0501024, Eur. Phys. J. {\bf A} (2005) 395.\vs
%%CITATION = HEP-LAT 0501024;%%

\bibitem{FMMS}
N.~Fettes, U.-G.~Mei{\ss}ner, M.~Moj\v zi\v s and S.~Steininger,
%``The chiral effective pion nucleon Lagrangian of order p**4,''
Annals Phys.\  {\bf 283} (2000) 273
[Erratum-ibid.\  {\bf 288} (2001) 249]
[arXiv:hep-ph/0001308].\vs
%%CITATION = HEP-PH 0001308;%%\

\bibitem{FMdel} 
N.~Fettes and Ulf-G.~Mei{\ss}ner,
%``Pion nucleon scattering in an effective chiral field theory with  explicit
%spin-3/2 fields,''
Nucl.\ Phys.\ A {\bf 679} (2001) 629
[arXiv:hep-ph/0006299].\vs
%%CITATION = HEP-PH 0006299;%%

\bibitem{BKMlec}
V.~Bernard, N.~Kaiser and Ulf-G.~Mei{\ss}ner,
%``Determination of the low-energy constants of the next-to-leading order
%chiral pion nucleon Lagrangian,''
Nucl.\ Phys.\ A {\bf 615} (1997) 483
[arXiv:hep-ph/9611253].\vs
%%CITATION = HEP-PH 9611253;%%


\bibitem{Hemmert}  T.~R.~Hemmert, Ph.D. thesis, University of Massachusetts at Amherst (1997) 
[UMI-98-09346-mc].\vs

\bibitem{Johnson} 
K.~Johnson and E.~C.~G.~Sudarshan,
%``Inconsistency Of The Local Field Theory Of Charged Spin 3/2 Particles,''
Annals Phys.\  {\bf 13} (1961) 126.\vs
%%CITATION = APNYA,13,126;%%

\bibitem{resonanceSSE}
V.~Bernard, T.~R.~Hemmert and Ulf-G.~Mei{\ss}ner, {\it forthcoming}.\vs

\bibitem{Dashen}
R.~F.~Dashen, E.~Jenkins and A.~V.~Manohar,
%``The 1/N(c) expansion for baryons,''
Phys.\ Rev.\ D {\bf 49} (1994) 4713 [Erratum-ibid.\ D {\bf 51} (1995) 2489]
[arXiv:hep-ph/9310379].\vs
%%CITATION = HEP-PH 9310379;%%

\bibitem{Vladimir}
V.~Pascalutsa and D.~R.~Phillips,
%``Effective theory of the Delta(1232) in Compton scattering off the
%nucleon,''
Phys.\ Rev.\ C {\bf 67} (2003) 055202 [arXiv:nucl-th/0212024].\vs
%%CITATION = NUCL-TH 0212024;%%

\bibitem{MZdel}
M.~R.~Schindler, J.~Gegelia and S.~Scherer,
%``Infrared regularization of baryon chiral perturbation theory
%reformulated,''
Phys.\ Lett.\ B {\bf 586} (2004) 258
[arXiv:hep-ph/0309005].\vs
%%CITATION = HEP-PH 0309005;%%

\bibitem{Fuhrer}A. Fuhrer, Diploma Thesis ``The
nucleon in finite volume", university of Berne (2004).\vs

\bibitem{HPW}
M.~Procura, T.~R.~Hemmert and W.~Weise,
%``Nucleon mass, sigma term and lattice QCD,''
Phys.\ Rev.\ D {\bf 69} (2004) 034505
[arXiv:hep-lat/0309020].\vs
%%CITATION = HEP-LAT 0309020;%%

\bibitem{young}
R.~D.~Young, D.~B.~Leinweber, A.~W.~Thomas and S.~V.~Wright,
%``Chiral analysis of quenched baryon masses,''
Phys.\ Rev.\ D {\bf 66} (2002) 094507 [arXiv:hep-lat/0205017].\vs
%%CITATION = HEP-LAT 0205017;%%

\bibitem{BuM}
P.~B\"uttiker and U.-G.~Mei{\ss}ner,
%``Pion nucleon scattering inside the Mandelstam triangle,''
Nucl.\ Phys.\ A {\bf 668} (2000) 97
[arXiv:hep-ph/9908247].\vs
%%CITATION = HEP-PH 9908247;%%


\bibitem{GLS}
J.~Gasser, H.~Leutwyler and M.~E.~Sainio,
%``Sigma Term Update,''
Phys.\ Lett.\ B {\bf 253} (1991) 252.\vs
%%CITATION = PHLTA,B253,252;%%




\end{thebibliography}
\end{document}